\documentclass{aa}

\begin{document}

\title{Leonid Electrophonic Bursters}

\author{Martin Beech\inst{1} \and Luigi Foschini\inst{2}}

\institute{Campion College and Department of Physics, The University 
of Regina, Regina, Saskatchewan, Canada S4S 0A2 (email: 
Martin.Beech@uregina.ca) 
\and Institute TeSRE -- CNR, Via Gobetti 101, 
I-40129 Bologna, Italy; (email: foschini@tesre.bo.cnr.it)}

\offprints{M. Beech}

\date{Received 4 October 2000 / Accepted 5 December 2000}

\abstract{We investigate the conditions under which Leonid meteoroids might
generate short duration (burster) electrophonic sounds.  A `first order'
theory is employed to estimate the approximate electron number density in the
meteoroid ablation column as a function of time.  Using the threshold
conditions discussed in an earlier communication (Beech \& Foschini~1999) we
find that Leonid meteoroids more massive than about 0.1 kg can potentially
generate short duration electrophonic bursters.
\keywords{Meteors, meteoroids}}

\maketitle

\section{Introduction}
The existence, or not of electrophonic sounds has long been a 
contentious issue within meteor studies.  The essential mystery of 
what are now called electrophonic sounds is that observers report 
hearing ``crackling'' or ``rushing'' and sometimes ``popping'' sounds 
at the same time as seeing a fireball.  The sounds often mimic the 
behaviour of the meteor, rising and falling as a meteor fragments and 
flares.  While such sounds have been reported throughout antiquity, 
there has until very recently been little experimental investigation 
of the phenomenon in the field (Beech et al.  1995; Beech \& Nikolova 
1999; Garaj et al.  1999).  For all the contention, however, Keay 
(1980, 1993) and later on Bronshten (1983), have presented a clearly described
and testable model for the generation of enduring electrophonic sounds.  In
addition, Beech \& Foschini (1999) have presented the outlines of a model for
the origin of the short duration, so called, burster electrophonic sounds.
Generally speaking enduring electrophonic sounds may be associated 
with the appearance of slow, sporadic fireballs (Keay \& Ceplecha 
1994), but electrophonic bursters have reportedly been heard during 
Leonid, and Lyrid outbursts and annual Perseid meteor shower displays.

In this paper, we are specifically concerned with the Leonid meteoroid
stream.  In an earlier study, Beech (1998) found that enduring 
electrophonic sounds might be produced from (of order of) metre--sized 
Leonid meteoroids.  Clearly, such large objects are not going to 
be abundant in the Leonid stream at any one time, but it has been 
suggested by Beech \& Nikolova (2000) that they might be deposited 
into the stream during mantle ejection events associated with the aging 
of comet 55P/Tempel--Tuttle.  Reports gathered by Olmsted during the 
1833 Leonid storm indicate that enduring electrophonic sounds, as well 
as electrophonic bursters, were heard and this is indicative of the
presence of at least a few large Leonid meteoroids.  Of more recent note, however, electrophonic bursters have reportedly been heard during both the 1998 
(Darren Talbot, personal communication) and 1999 (Drummond et al., 
2000; Beatty, 2000) Leonid outbursts.

\section{The electrophonic burster model}
The various catalogues that have been compiled on electrophonic sounds 
indicate that about 10\% of the reports fall into what we 
call the ``burster'' category.  The bursters are characterized by 
being short in duration (typically a second or less), and they are 
often described as sounding like ``pops'', or ``vits'' and even 
``sharp clicks''.  The other key characteristic of the burster events 
is that they are often, but not always, associated with a single, 
dramatic break--up event of the ablating meteoroid.

Beech \& Foschini (1999) have suggested that the burster phenomenon 
can be explained in terms of a shock propagating within a meteor's 
plasma column.  Within the framework of this model, it is argued, that the rapid movement of the electrons, with respect to the much more massive and slower moving ions, generates a sizable space charge (see e.g., Zel'dovich \& Raizer, 1967). A transient electrical pulse is generated in response to the development of the space charge, and provided the resultant electrical field strength variations are large enough it is suggested, following Keay ~(1980), that they might trigger the generation of audible sounds through an observer localized transduction process. The shock wave is produced, Beech \& Foschini (1999) suggest, during the catastrophic break--up of the parent meteoroid. Under ideal conditions the electric field generated would be symmetric, propagating equally in all directions.  However, as is more likely, the 
presence of fluid dynamic instabilities will lead to preferential propagation in certain directions.  Specifically, since the meteor plasma can be thought of as a fluid with a higher density than the surrounding atmosphere, both Rayleigh--Taylor and Kelvin--Helmholtz instabilities might develop.  In the possible presence of these instabilities, any shock front will soon become distorted.  One may therefore envision the situation in which the electric field is either focused or defocused in random directions by the growing filaments and small scale perturbations in the shock. Gull~(1975) has described the complex instability modification of a shock front within the context of a stellar supernovae model.  The same processes are likely to occur in airburst explosions, with obvious changes in energy scale.  In addition to the presence or not of appropriate transduction material, this instability feature may explain why electrophonic sounds often appear to be highly localized.

Experiments conducted by Keay \& Ostwald (1991) suggest that an 
electric field strength of at least 160~V/m is required to generate 
electrophonic sounds.  Beech \& Foschini (1999) find that such a 
threshold electric field can be generated provided the electron number 
density within the plasma exceeds $n_{\mathrm{e}}\approx 4\cdot
10^{18}$~m$^{-3}$.

\begin{table*}[!t]
\centering
\caption{Critical Leonid meteoroid mass limits for generating
electrophonic bursters.  Column one indicates which form of the ionization coefficient was employed in the calculations; the second column indicates the meteoroid composition; column
three corresponds to the assumed zenith
angle of entry; column four indicates the height at which $n_{\mathrm{e}}$
first exceeds the $10^{19}$~m$^{-3}$ threshold; column five indicates
the approximate visual magnitude of the meteoroid at $h_{\mathrm{crit}}$.}
\begin{tabular}{llcccc}
\hline
Ionization Theory & Composition & $Z$ [deg.]  & $m_{\mathrm{crit}}$ [kg] &
$h_{\mathrm{crit}}$ [km] & $M_{\mathrm{vis}}$\\
\hline
Jones (1997) & Type IIIA & 0 & 0.028 & 80.35 & $-8.6$\\
{}           & {}        &20 & 0.031 & 80.60 & $-8.7$\\
{}           & {}        &45 & 0.051 & 81.30 & $-8.9$\\
{}           & {}        &60 & 0.095 & 82.23 & $-9.2$\\
\hline
{}           & Type IIIB & 0 & 0.12 & 85.80 & $-10.4$\\
{}           & {}        &20 & 0.14 & 85.90 & $-10.4$\\
{}           & {}        &45 & 0.22 & 86.73 & $-10.7$\\
{}           & {}        &60 & 0.41 & 87.60 & $-11.0$\\
\hline
Bronshten (1983) & Type IIIA & 0  & 0.005 & 84.10 & $-6.8$\\
{}               & {}        &20 & 0.006 & 84.25 & $-6.8$\\
{}               & {}        &45 & 0.009 & 84.93 & $-7.0$\\
{}               & {}        &60 & 0.017 & 85.85 & $-7.3$\\
\hline
{}               & Type IIIB &0  & 0.021 & 89.30 & $-8.5$\\
{}               & {}        &20 & 0.023 & 89.43 & $-8.6$\\
{}               & {}        &45 & 0.039 & 90.20 & $-8.8$\\
{}               & {}        &60 & 0.070 & 91.08 & $-9.1$\\
\hline
\end{tabular}
\label{TAB1}
\end{table*}

\section{The ionization coefficient}
In this section we shall attempt to constrain the critical mass above 
which a Leonid meteoroid might produce an electrophonic burster.  We 
shall employ order of magnitude arguments at this stage, leaving 
refinements of the physics to future detailed numerical modeling.  The 
great complexity of the meteoroid interaction with the Earth's 
atmosphere is discussed, for example, in Ceplecha et al.~(1998) and
Foschini~(1999).

The variation in the electron line density, $q$, produced by an 
ablating meteoroid can be described as (Kaiser 1953; Hughes 1978):

\begin{equation}
	q=-\frac{\beta}{\mu V}\frac{dm}{dt}
	\label{e:eld1}
\end{equation}

\noindent where $V$ is the velocity, $\beta$ is the ionization 
coefficient, $\mu$ is the mean atomic mass of meteoroid atoms, and 
$dm/dt$ is the mass ablation rate.  The mass loss rate can be 
derived by solving the standard single--body equations of meteoroid 
ablation (see, e.g.  \"{O}pik 1958, McKinley 1961).

To first order, the electron line density may be converted to an 
electron number density $n_{\mathrm{e}}$ via the relation:

\begin{equation}
	q=\pi r_{\mathrm{i}}^{2}n_{\mathrm{e}}
	\label{e:conv}
\end{equation}

\noindent where $r_{\mathrm{i}}$ is the initial train radius.
Eq.~(\ref{e:conv}) is the limiting initial case of the diffusion 
equation when $t\approx 0$ (see e.g., Mckinley, 1961).  Provided one 
can establish analytically reasonable expressions for the ionization 
coefficient, $\beta$, and the initial train radius, $r_{\mathrm{i}}$, then
Eqs.~(\ref{e:eld1}) and (\ref{e:conv}) may be solved for numerically.  
In this manner, the characteristics of the Leonid meteoroid just 
capable of producing an electron number density of, say, 
$10^{19}$~m$^{-3}$ (this places $n_{\mathrm{e}}$ above the burster threshold
condition derived by Beech \& Foschini 1999) can be determined.

The core of the problem is the determination of the ionization 
coefficient $\beta$.  It is, by definition, the ratio of the number of 
free electrons produced to the number of meteoroid atoms vaporized.  
The ionization coefficient is strongly dependent upon the relative 
speed with which the meteoroid's ablated atoms and atmosphere molecules
collide. Typically the ionization coefficient is expressed as a power law
in the velocity

\begin{equation}
	\beta=\beta_{0}\cdot V^{n}
	\label{e:beta1}
\end{equation}

\noindent where $\beta_{0}$ and $n$ are coefficient to be calculated.

Massey \& Sida~(1955) and later on Sida~(1969) studied the collision
processes in meteor trails and argued that $\beta$ can be expressed as the
ratio of the ionizing and momentum--loss cross sections. It is the sensitivity
of the ionization cross section to the relative velocity that introduces the
velocity dependency in $\beta$. Massey and Sida proposed several different
ways of weighting the cross sections, taking into account different factors
such as the angle of scattering (see Jones, ~1997 for a discussion of the
inherent problems with the Massey and Sida model).  Indeed, even though the
definition of $\beta$ is quite straightforward, the complexity of particle
dynamics in the meteor plasma complicates the situation greatly, and the
correct weighting factors are still a topic of debate (for a recent review on
particle dynamics in meteor plasma's see Dressler \& Murad 2000).

Bronshten~(1983) presents a detailed analysis of the empirical and
theoretical calculations of $\beta$ and finds that:

\begin{equation} 	
\beta=5.47\cdot 10^{-7}\cdot V^{3.42} 	
\label{e:beta2}
\end{equation}

where the velocity V is expressed in km/s. Jones~(1997) has recently provided a thorough review of the methods by
which $\beta$ might be calculated, and finds that for visual meteors
with velocities in the range of $30$ to $60$~km/s.

\begin{equation}
	\beta=4.91\cdot 10^{-6}\cdot V^{2.25}
	\label{e:beta}
\end{equation}

In the theory developed by Jones $\beta < 1$, while Bronshten considered
the situation when $\beta$ is greater than unity.  In this later case,
secondary ionization is included under the supposition that at very high
relative velocities a single meteoroid atom can produce more than one free
electron. Within the context of this first order study we choose to use both
approximations for the ionization coefficient and present two sets of
comparative calculations.

As with the ionization coefficient, the initial train radius is also a
difficult term to quantify.  Generally speaking it will be of order the
atmospheric mean free collision length.  Radar measurements of meteor trains
have been used to determine the initial train radius as a function of
atmospheric height.  Baggley (1970) found, for example, that the initial train
radius was some 3~m at an altitude of 115 km, and some 0.5~m at 90 km
altitude.  These values of the initial train radius are some 2 to 20 times the
mean free path lengths at the respective atmospheric heights.  Jones (1995)
has derived a relationship for the initial train radius as a function of the
number density $n_{\mathrm{a}}$ of air molecules.  He finds:

\begin{equation}
	r_{\mathrm{i}}=2.845\cdot 10^{18}\frac{V^{0.8}}{n_{\mathrm{a}}}
	\label{e:inir}
\end{equation}

\noindent where $V$ is the velocity measured in km/s.  The power of 
$0.8$ in the velocity term describes the energy sensitivity of the 
ionic scattering cross-section.

\section{The critical mass}
Gathering together Eqs.~(\ref{e:eld1}) through (\ref{e:inir}) and 
solving them in conjunction with the single body ablation equations we 
may proceed to determine the electron number density as a function of 
atmospheric height for a given initial mass.  The initial velocity for 
Leonid meteoroids is determined from the observations to be 71 km/s.  The 
choice of an appropriate meteoroid composition is rather problematic, 
but we shall consider the so--called Type IIIA cometary and Type IIIB 
soft cometary ablation parameters (see, Ceplecha et al.~1998).  When 
solving the single body ablation equations we use the density--height 
profile provided by the MSIS--E--90 Earth atmosphere model.

The critical mass, $m_{\mathrm{crit}}$, that we are looking for determines the
mass of  the Leonid meteoroid that will just generate a maximum electron
number density of $10^{19}$~m$^{-3}$ somewhere along its ablation track.
Meteoroids more massive than $m_{\mathrm{crit}}$ will produce electron number
densities higher than the critical value, and subsequently should such 
meteoroids catastrophically disintegrate (when 
$n_{\mathrm{e}}>10^{19}$~m$^{-3}$) then the burster mechanism described by
Beech \& Foschini (1999) may come into play.

The results of our calculations are presented in Tab.~\ref{TAB1}.
The calculations begin at an atmospheric height of 190~km
and various zenith angles ($Z$) of meteoroid entry have been assumed. We have used a constant luminous efficiency of 1

The data gathered together in Tab.~\ref{TAB1} suggest
that the minimum mass for a Leonid meteoroid to produce an 
electrophonic burster is somewhere in the range of 5 to 400~g depending
upon zenith angle, composition and the velocity sensitivity of the
ionization coefficient.  The predicted brightness of the burster meteors
is in the magnitude range of $-7$ to $-11$.

For a given zenith angle the variation in composition (Type IIIA or IIIB)
results in about a factor of 4 variation in the critical mass estimate.
For a given composition and zenith angle, the uncertainty in the ionization
parameters (expressed through the use of Bronsthen's formula and that of
Jones) results in a variation of about 6 in the critical mass. 

\section{Some comments on observational techniques}
The contentious aspect of the debate surrounding electrophonic sound
production has mostly focused on the issue of observational reliability.
Eye--witness and anecdotal accounts are of little value, beyond bolstering
the cause, in the modern era. What is required is unambiguous instrumental
data. Literally, data that can stand--up to theoretical analysis and
detailed scrutiny. Beech et al.~(1995) have reported on the detection
of an unusual VLF transient associated with the disruption of a $-10$
visual magnitude Perseid. More recently Garaj et al.~(1999) have claimed
an associative link between a series of very short--duration very low frequency radiation (VLF) transients
and a magnitude $-8$ Leonid fireball. An as yet to be published study
conducted by Price et al during the 1999 Leonid storm (see Beatty~2000)
also claims to have recorded many meteor -- VLF transient coincidences.

It is our basic belief that none of the observational studies
published to date can claim to have clearly established a casual
linkage between the generation of a characteristic VLF transient
signal and the passage of a fireball. We offer this statement in
the sense that none of the published papers present instrumental
data that unambiguously shows the coincident detection of a meteor
and VLF or electrophonic sound transient.

While it is not our intention to be overly critical of honest effort,
we would also like to point out that it is not obviously clear that the
correct line of experimental attack has always been employed. Firstly,
we should ask ourselves what it is that we are actually trying to measure.
Electric and magnetic fields behave differently depending on
the distance $R$ between the source and the point of measurement (see, for
example, Jackson~1975, chapter~9). Indeed, we can distinguish three
characteristic zones: the near field, the far field, and the intermediate
zone. Comparing the distance $R$ to the source with the wavelength
$\lambda$, we have

\begin{itemize}
\item near field  : $R << \lambda$
\item far field   : $R >> \lambda$
\item intermediate: $R \approx \lambda$
\end{itemize}

Audible frequencies fall in the range of $0.02-4$~kHz, therefore the
minimum wavelength is 75~km, this, we note, is comparable to the meteor
heights displayed in Tab.~1. This is valid if the mundane objects, acting as
signal transductors, do not appreciably change the frequency of the incoming
electromagnetic radiation. If the transductors do induce a frequency shift,
it is possible that the incoming radiation has a higher frequency
(and thus, a smaller wavelength).

For more massive meteoroids that penetrate deep into the Earth's atmosphere,
we are essentially in the intermediate and near field regions. Only in the far
field zone would we actually have a plane transverse electromagnetic wave (and
therefore a radiation field). In the burster model discussed above, it is most
probable that the intermediate field condition applies and hence the electric
and magnetic fields will behave differently. Within this context it may be no
coincidence that the VLF transient events reported by Beech et al.~(1995) and
Garaj et al.~(1999) relate to meteors seen close to the horizon (but see
below). Indeed, if the source is close to the horizon the distance from the
observer can be many hundreds of kilometres. In the long--range case, we have
a radiation field and consequently the transmission of energy via the Poynting
vector. Before embarking on an experiment to study electrophonic transients a
clear decision has to be made with respect to whether it is the electric or
the magnetic field component that is to be measured. This decision in turn
dictates the choice of antenna to be used in the experiment. The Keay and
Bronshten electrophonic model is based upon the relaxation of the geomagnetic
field, approximated as a magnetic dipole source, and to study this component
one must employ a loop antenna valid for frequencies in the range 50~Hz --
50~kHz. The mechanism outlined by Beech and Foschini for burster
electrophonics may be thought of as generating an electric dipole source, and
consequently to study this component a vertical wire antenna should be used.
If one wishes to test the transduction process directly a well--calibrated
detector and microphone system will have to be developed.

In addition to the appropriate selection of an antenna, care should also be
directed towards the choice of signal analyzer. The best choice of detector is
probably an Electro Magnetic Interference (EMI) analyzer. Such devices are
typically sensitive over a wide dynamical range of frequencies. And indeed, we
note, that since one is trying to measure implusive or rapidly changing
broadband signals an instrument with a small dynamical range will seriously
compromise the quality of measurements being made.

The model for electrophonic sound generation described by Keay ~(1980; 1993)
makes a clear prediction. That is, the VLF signal generated by the meteoroid
interaction with the geomagnetic field will be sustained, possibly observable over an extended
period of time, and it will be distinct from background atmospheric sources. While
the exact VLF signal characteristics can not be predicted at this stage, it is
unlikely that it constitutes a series of very--short duration transients. In
this respect, we do not at this stage fully accept that the very short
duration VLF transient signals presented by Garaj et al.~(1999) in their
survey paper represent anything other than background, atmospheric events.
They may represent a real meteor related signal, but the evidence it not
compelling. The VLF transient signal presented by Beech et al. ~(1995), on the
other hand, has characteristics (signal duration and distinctness from natural
atmospheric sources) that make it more believable as an actual meteor related
event. Our point here is that unless the meteor generated VLF signal has
characteristics that clearly distinguishes it from background sources, then
one cannot simply claim that because short-lived VLF transient are observed at
the same time that a meteor is seen that the two observations are causally
related. It is not clear to us as yet whether the electrophonic bursters are
likely to generate a significant VLF signal.

\section{Discussion and conclusions}
Our estimate of the minimum Leonid meteoroid mass for generating electrophonic
burster sounds is much smaller than that derived for long enduring
electrophonic sounds (of order a hundred grams rather than 800~kg).  This
result, in fact, sits well with the observations since we can be reasonably
sure that of order one hundred gram and larger mass meteoroids do exist within
the Leonid stream (see e.g., Spurny et al.~2000 and Bellot Rubio et al.,~2000).

We re--iterate, however, that we do not expect every Leonid fireball of
magnitude $-7$ (and brighter) to generate electrophonic burster sounds.  The
key point is that the meteoroid has to develop a shock wave at the same time
that the electron number density in the plasma column exceeds
$10^{19}$~m$^{-3}$. In addition, to generate a measurable signal in the VLF
range the meteor must be placed at a distance greater than some 75~km from the
observer.

Many of the Leonid fireballs recorded in 1998 where found to began
rapid ablation at heights in excess of 120 km. In addition, the average end
height of Leonid fireballs was found to be about 85 km (Spurny et al.~2000).
These results certainly indicate that Leonid meteoroids are made of very
friable and easily ablated material. Indeed, Spurny et al.~(2000) classified
all but one of the Leonid fireballs that they recorded as type IIIB, which is
typical of the weakest interplanetary bodies. The model outlined above suggests that a meteoroid has to penetrate to an
altitude of about 80 to 90 km before it undergoes catastrophic disruption.  On
this basis we see no inherent reason why some Leonid fireballs might not
produce electrophonic bursters.

One of the key aspects of the enduring
electrophonic sound model developed by Keay and Bronshten was the production
of very low frequency (VLF) radiation.  This radiation is generated through an
interaction of the highly ionized and turbulent meteor plasma train with the
Earth's magnetic field.  It is the transduction of the VLF radiation by
objects close to the observer that results in the generation of audible
sounds.  In contrast, the electrophonic burster model outlined by Beech and
Foschini does not specifically predict the generation of any VLF signal.
Rather, it predicts the generation of a short--lived transient pulse (or
pulses) in electric field strength. The observational campaigns conducted to
date have mostly focused upon the detection of VLF radiation transients
through monitoring magnetic field variations. Electrophonic bursters, we
argue, are more likely to be observed through the generation of electric field
transients.

\begin{acknowledgements}
We extend our appreciation to the referee, L.~Bellot Rubio, for his comments
and suggestions. This work has been partially supported by a grant from the
Natural Sciences and Engineering Research Council of Canada and partially by
MURST Cofinanziamento 2000. This research has made use of \emph{NASA's
Astrophysics Data System Abstract Service}.
\end{acknowledgements}

\end{document}